\documentclass[11pt,a4paper]{article}
\usepackage{amsmath,amssymb,amsfonts}
\usepackage{graphicx}
\usepackage{hyperref}
\usepackage{geometry}
\title{Indirect Computing Model with Indirect Formal Method}
\author{Xiaohui Zou$^{1,2,3}$, Shunpeng Zou$^{1}$\\
$^{1}$Institute for Higher Education, China University of Geosciences (Beijing), Beijing, China\\
$^{2}$Institute of Synergistic Cultural Gene Engineering, Tsinghua Science Park, Zhuhai, China\\
$^{3}$Sino-US Project: UC Berkeley Searle Research Bilingual Information Processing Group\\
\texttt{zouxiaohui@pku.org.cn}}
\date{}

\begin{document}

\maketitle

\begin{abstract}
This paper, from the perspective of a collaborative intelligent computing system formed by combining human-computer interface and collaborative computing programs, discusses the principles of optimized cloud computing technology supported by the combination of an indirect computing model and an indirect formal method. On the basis of systematically reviewing the influence of previous theoretical achievements (Turing's computability theory, Kleene's formal theory of small strings, von Neumann's digital computer architecture, and Turing's hypothesis on AI judgment) on the mainstream general-purpose digital computer paradigm, the author focuses on introducing an indirect computing model and an indirect formal theory compatible with both large and small strings. Using Chinese information data as an example, the design concept of a collaborative intelligent computing system prototype is presented. The significance is that this achievement facilitates optimization of cloud computing from data centers to knowledge centers.
\end{abstract}

\subsection*{Keywords}
Computing theory; formal theory; human-computer interaction; collaborative computing

\section{Introduction}

This paper, from the perspective of a collaborative intelligent computing system formed by combining human-computer interface and collaborative computing programs, discusses the principles of optimized cloud computing technology supported by the combination of an indirect computing model and an indirect formal method. Previous theoretical achievements — Turing's computability theory \cite{turing1936}, Kleene's formal theory of strings \cite{kleene}, von Neumann's digital computer architecture \cite{vonneumann}, and Turing's hypothesis on AI judgment \cite{turing1950} — are the foundation of this research. The indirect computing model provides a series of good algorithms; the indirect formal method provides optimized data structures; their sum equals the twin Turing machine virtual computing program. First, it converges Turing's computable numbers in the target domain. Furthermore, it reveals the transformation constraints involved in the double transformation process of computational complexity NP-complete problems \cite{np}: how to deepen understanding from P to NP and how to simplify expression from NP to P. Then, by adopting an indirect formal path compatible with both large and small strings, it not only improves Kleene's theory of small strings but also extends beyond string processing to forms such as characters, formulas, diagrams, tables, sounds, images, three-dimensional objects, and living bodies. The basic design concept of a collaborative intelligent computing system \cite{zou2008} is introduced using Chinese information data as an example. The significance is that this achievement facilitates optimization of existing cloud computing \cite{cloud} from data centers \cite{datacenter} to knowledge centers \cite{knowledgecenter}.

\section{Exploration from Macro, Micro, and Meso Perspectives}

Genetic text and its systems engineering blueprint, ideal classification sets, and basic information laws are explored from macro, micro, and meso perspectives respectively, concerning three aspects: genetic text and its possible human-computer interface forms, data structures and their possible mechanisms of division/combination, and formal information and its inherent principles or possible transformation rules. Thus, the cognitive premise or thinking paradigm of this research is established.

\subsection{Describing Genetic Text and Its Systems Engineering Blueprint Using Chinese Characters as an Example}

This paper restricts discussion to genetic text and its systems engineering blueprint \cite{xmind1} — a scope that is both indirectly computable and indirectly formalizable, thereby more amenable to natural language processing and understanding, and concerning genetic text and its possible human-computer interface forms.
\begin{figure}
    \centering
    \includegraphics[width=1\linewidth]{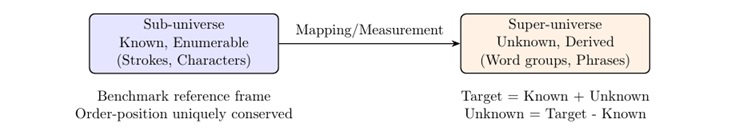}
    \caption{Blueprint of cultural gene systems engineering matched with optimized data structures and good algorithms}
    \label{fig:placeholder}
\end{figure}
As shown in Figure 1, Chinese characters exhibit two types of genetic text — basic strokes of Chinese characters and monosyllabic Chinese characters, which can be regarded as elements of the sub-universe. Their characteristic is that they are enumerable and easy to enumerate; these are simple basic texts. The second type includes Chinese characters and their radicals (as a whole), and the \textit{yan} (characters) and \textit{yu} (character groups at various levels) of Chinese — that is, complex derived texts. Their characteristic is that, after ideal classification, they can be placed in twin Turing machine double lists, making targeted search possible.

Why can the author assert this? Because the blueprint of cultural gene systems engineering (Figure 1) tells us several truths. It reveals an optimized data structure expressed as:
\[
\text{Total text universe} = \text{Sub-universe} + \text{Super-universe}
\]
which also means
\[
\text{Target domain} = \text{Known domain} + \text{Unknown domain}
\]
and the convergence strategy leading to good algorithms:
\[
\text{Unknown domain} = \text{Target domain} - \text{Known domain}
\]
The first formula describes an ideal classification set; the latter two formulate general and narrow information equations.

From \{basic strokes\}, we can construct \{various radicals\}, and then \{monosyllabic characters\}; from \{monosyllabic characters\} we can construct \{character groups at various levels\}. These two types of Chinese text genes and their construction principles, together with the ideal classification set (Figure 2), basic information laws (Figure 3), and twin Turing machines (Figure 4), yield a collaborative intelligent computing system strategy for efficiently processing the genetic and derived texts shown in Figure 1.

\subsection{Describing Ideal Classification Sets Using Binary as an Example}

\begin{figure}
    \centering
    \includegraphics[width=1\linewidth]{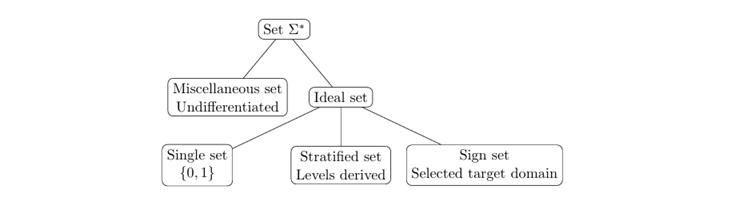}
    \caption{Diagrammatic illustration of ideal classification sets and their possible mechanisms of division/combination}
    \label{fig:placeholder}
\end{figure}
As shown in Figure 2, the author reclassifies \(\Sigma^{*} = \{\varepsilon, 0, 1, 00, 01, 10, 11, 000, 001, 010, 011, \dots\}\) into: single sets, stratified sets, sign sets, and the original undifferentiated miscellaneous set \((\Sigma^{*})\). Using binary numbers to describe ideal classification sets not only reveals the basic information propositions embedded within, but also provides a feasible strategy for further optimizing various data structures. More importantly, this exposition is concise and effective for identification, understanding, and expression.

\textbf{Principle 1:} Optimizing ideal classification sets further distinguishes ordinary sets into two major types: miscellaneous sets and ideal sets. Among them, the former includes Cantor sets and generalized sets (populations, covering natural distributions, self-organizing groups, and other-organized groups); the latter includes single sets, stratified sets, and sign sets. Optimizable ideal sets are a focus of this paper.

\textbf{Definition 1 (Single Set):} A single set contains only one type of element. For binary numbers: \(\{0,1\}\). For decimal numbers: \(\{0,1,2,3,4,5,6,7,8,9\}\). Tuples derived by copying and recombining these elements do not exceed the range of possible permutations of the genetic text symbols. They only cause structural changes through duplication and recombination, and only increase total quantity or transform basic order-position relationships. Hence, it is named sub-universe. The characteristic is that elements of a single set or sub-universe are fixed and non-repetitive; the order and position (abbreviated as order-position relations) of elements within its scope are uniquely conserved.

\subsection{Describing Basic Information Propositions Using Ideal Classification Sets}

\subsubsection{First Basic Information Proposition}

\textbf{Theorem 1:} For sub-universe elements, \textit{order-position relations are uniquely conserved}. The sub-universe can serve as a reference frame for measuring and computing the super-universe, i.e., a single set can norm the evolutionary development of tuples in stratified sets.

\textbf{Definition 2 (Stratified Set):} A stratified set is derived stepwise from the above single set through replication and recombination. It determines the possible tuples of elements at each level, characterized by evolutionary layering. For binary numbers: \{0,1\}, \{00,01,10,11\}, \{000,001,010,011,110,101,100,111\}, \ldots. The evolutionary ladder includes the first level directly copied from the single set, then subsequent levels derived by continuous copying and recombination. Its characteristic is that both the stratified set and its tuples have non-repeating order-position relations; each level and tuple has a fixed quantity and unique order-position relations. Thus, each level can serve as a response reference frame for enumerating or optimally searching the contained elements/tuples.

\textbf{Definition 3 (Sign Set):} A sign set is a scope or domain selected by human subjects or computer agents from single and stratified sets based on specific goals, also called the target domain. It is a set composed of specific single sets and specific levels of stratified sets. For example, the collaborative intelligent computing system described in this paper selects \{binary numbers\}, \{decimal numbers\}, \{English letters\}, \{Chinese strokes\} and various sign sets built upon them or equivalent to them, such as a fixed number of \{monosyllabic characters\} familiar to students of a certain age, and layered \{Chinese character groups\} recording \{common knowledge\} and \{specialized knowledge\} — equivalent to corresponding \{English words and phrases\}. The characteristic is that the reference frame (benchmark) involved in the sign set/target domain is fixed, and the levels of the response reference frame are determinable. Although tuples in the unknown domain are uncertain, those in the known domain are determinate, making it easy to select optimized data structures and good algorithms, thereby significantly accelerating the enumeration and search of tuples at each level within the target domain. This benefits the bilingual/biliterate processing of Chinese and optimizes new strategies for statistical machine translation.

\subsubsection{Second Basic Information Proposition}

\textbf{Theorem 2:} For tuples of the super-universe (including sub-universe elements), once two sequences of corresponding data appear in a specific target domain (whether known or unknown), as long as they can be placed in a relation of \textit{synonymy parallelism} (including agreement parallelism) and satisfy a one-to-one functional correspondence, they can be mutually transformed or substituted between the two sequences under pre-appointed rules.
\begin{figure}
    \centering
    \includegraphics[width=1\linewidth]{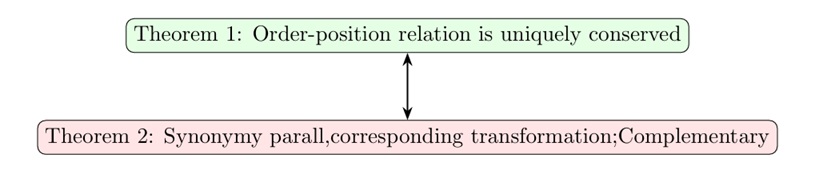}
    \caption{Diagram of basic information propositions}
    \label{fig:placeholder}
\end{figure}

As shown in Figure 3, both the description of sub-universe elements and super-universe tuples satisfy the first basic information proposition (unique conservation of order-position relations) with the feature of "different-arrangement order-interest simplicity-beauty". Compared with the second basic information proposition, they belong to two levels: the first is most fundamental. The reason the second proposition occupies a prominent position among the three supporting rules shown in Figure 3 is its universality and key role. The second basic information proposition is the basic rule for mutual transformation, abbreviated as the "synonymy parallelism, corresponding transformation" rule. Algebraic identity transformations rely on this rule; the basic formula of generative grammar (S = NP + VP) also relies on it. Many other examples satisfy the second basic information proposition. If single sets and stratified sets together mainly determine uniquely conserved order-position relations — i.e., authentic information recorded by genetic texts that can be evaluated by truth values for sub-universe elements and super-universe tuples — then such authentic information is appropriate and understandable. Whether for the undifferentiated miscellaneous set or for the finely partitioned sign set, in essence, only through the above single and stratified sets can we achieve more thorough understanding and more appropriate expression.

\section{Combination of Indirect Computing Model and Indirect Formal Method}

Because the indirect computing model has the basic features of distributed and parallel computing that can be both separated and combined between computer agents and human subjects, together with its supporting indirect formal method, it forms a twin Turing machine resembling a balance. The left list (data structure type composed of natural number sequential codes) is like customized standardized weights; the one-to-one corresponding right list (data structure type with reserved slot positions) is like arbitrary individual items to be weighed. The difference is that the balance, weights, and "items" are virtual. Therefore, we can call the second basic information proposition the \textbf{balance rule}. Hence, the second basic information proposition is the theoretical basis for constructing the twin Turing machine.

\begin{figure}
    \centering
    \includegraphics[width=1\linewidth]{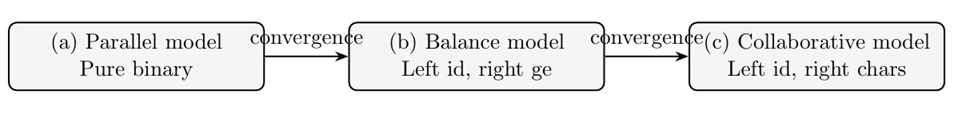}
    \caption{Schematic principle of twin Turing machines gradually converging from (a) to (c)}
    \label{fig:placeholder}
\end{figure}
As shown in Figure 4, the left (a) is a virtual twin Turing machine composed of two parallel Turing machines; the middle (b) and right (c) are equivalent forms, each with its own characteristics. (b) describes a balance-type measurement conversion device built on the "synonymy parallelism, corresponding transformation" rule; its construction principle is illustrated by embodiment (c)\cite{xmind1}, based on the finite set of textual symbols formed by enumerable Chinese monosyllabic characters. The double list constructed according to Theorem 2, after ideal classification of generalized bilingual texts, enables a reasonable division of labor and high-level collaboration between numbers and characters or between machines and humans, thus featuring separable and combinable collaborative intelligent computing. Its operation achieves the usage effect of integrating standardization and personalization. The three basic forms (a, b, c) share a common feature: they all consist of left and right symmetrical virtual tables (VT\&L and VT\&R). Each also has its uniqueness:

\begin{itemize}
\item \textbf{(a) Parallel computing model}: the upper bound of computable numbers set by pure binary numbers.
\item \textbf{(b) Distributed computing model}: embodies the one-to-one correspondence between the left list decimal numbers and the right list computable grids, performing virtual computation in the form of virtual ge grids, thus providing a series of general conversion platforms for collaborative processing of different data structure types.
\item \textbf{(c) Virtual cloud computing model}: embodies the one-to-one correspondence between the left list decimal numbers and the right list indirectly computable monosyllabic characters (using Chinese as an example). It is the embodiment of the combination of the indirect computing model and indirect formal method.
\end{itemize}

This is the principle of twin Turing machines gradually converging from (a) to (c). The purpose is to provide a collaborative intelligent computing system platform for selecting optimized data structures and good algorithms. Because the computability upper bound converges successively from (a) to (b) to (c), and most importantly, the combination of the two virtual twin Turing machines (b) and (c) not only gives good convergence for indirectly computed textual data, but also provides typical embodiments for algorithm selection, data structure optimization, and ideal classification of generalized texts. Therefore, the twin Turing machine formed by combining the indirect computing model and indirect formal method is the modeling of Theorem 2. Among them, model (b) is an abstract, universal macro model, while model (c) is a practical micro model using Chinese as an example.

\textbf{Lemma 1 (Formalization of the second basic information proposition: information equation of the target domain):} When both the benchmark reference frame and the response reference frame are determined, because the target domain = known domain + unknown domain, we have \textit{unknown domain = target domain - known domain}, which at least has a distributed solution. The tuples thus indicated have order-position relations in the response reference frame that are easy to enumerate or search. Actual results can be strictly tested and measured by recall, precision, reuse rate, and reproducibility. The information equation of the target domain can be regarded as the formalized second basic information proposition, because it is essentially a transformation of the balance rule within a finite target domain — i.e., finding a solution path to the equation under the condition of a predictable result (an identity). The mathematical expression \(I_U = I_D - I_K\) is the narrow information equation of the target domain.

\section{Collaborative Intelligent Computing System Described Using Chinese Processing as an Example}

The author now presents a typical embodiment of a practical virtual twin Turing machine, using the optimization of Chinese information data structures as an example. This is a typical collaborative intelligent computing system. Its foundation and core is a database constructed by the author based on the actual needs of Chinese information processing, which is exactly a collaborative intelligent computing system based on the combination of the indirect computing model and indirect formal method. Because a well-ordered data structure naturally contains good algorithms, the monosyllabic characters of Chinese and the character groups derived from them can, on one hand, exploit the advantages of computer processing of standardized formal information automatically generated; on the other hand, give play to humans' familiarity with processing personalized formal information, allowing selection of commonly recognized phrases as usage examples explaining the specific meanings of each monosyllabic character — thus serving as the basis for further automatic computation and statistics by computer-aided systems.

\begin{figure}
    \centering
    \includegraphics[width=1\linewidth]{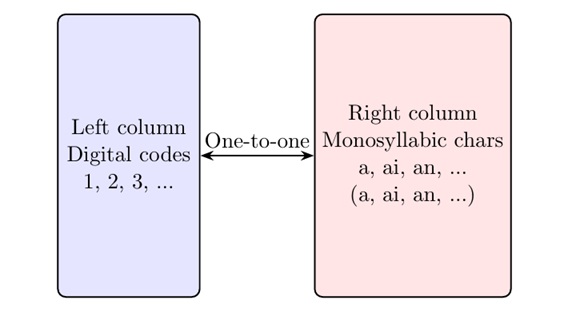}
    \caption{A relation database of \textit{yan} (char) and \textit{yu} (group) that has been indirectly formalized}
    \label{fig:placeholder}
\end{figure}
As shown in Figure 5, on the left are a series of parallel virtual twin Turing machines; in the middle are an easily computable matrix \((m \times n)\) and linear equations \((\sum a_{mn}x_n = b_m)\); on the right is the already indirectly formalized relation between \textit{yan} (characters) and \textit{yu} (word groups)\cite{xmind5}. Here, \textit{yu} (word groups) — i.e., phrases (equivalent to English words and phrases) — are derived level by level from \textit{yan} (characters). The distribution function formula \(A = (\sum n_i x_i)\) \cite{zhang2003} involves the generalized set (\(A\)) and sign sets (\(n_i x_i\)). For example, \{monosyllabic characters\}, by querying the single set \{basic strokes\} to form the stratified set \{radicals\}, can obtain and statistically count specific sign values and individual numbers in the target domain through a computer-aided system.

A detailed explanation of this example of a speech–language relation database is given below, in conjunction with Figure 5.

First, in the database shown in Figure 5, the target domain not only has realized the indirect formalization of \textit{yan} (characters) and \textit{yu} (word groups) and their interrelations, but also allows indirect computation (enumeration or search) at any time, which is very convenient, accurate, and efficient. In other words, the collaborative intelligent computing system represented by Figure 5 guarantees both recall and precision, and also allows very convenient, accurate, and efficient calculation and statistics of reuse rates. This is both an application example and a validation embodiment of Lemma 1. Specific operations can be performed in the target domain.

Second, it must be pointed out that for computers, this database is a series of standardized twin Turing machines; for humans, it is mainly a digital computer-aided tool platform. The reason it is special and can be called a collaborative intelligent computing system is mainly that its core involves a series of virtual twin Turing machines that embody the essence of collaborative intelligent computing. Guided by Theorem 1, Definition 1, Theorem 2, Definition 2, Definition 3, and Lemma 1, the three types of finely classified optimized data structures usually hidden in miscellaneous sets, as well as the good algorithms already discovered by mathematicians hidden in the left lists, can be transformed through the virtual twin Turing machine principle from "non-obvious" (NP problem being a special case) to "obvious" (P problem being a special case). Thus, the significance of the author's indirect formalization of the relation between Chinese \textit{yan} (characters) and \textit{yu} (word groups) can be gradually understood by colleagues accustomed to directly using various programming languages based on small character sets or small strings in an attempt to directly perform word segmentation on Chinese.

Finally, two points are worth mentioning:
\begin{enumerate}
\item According to both general linguistics and computational linguistics using Chinese as an example, it can be concluded that using the "character-based" (Yan-based) linguistic principle to process large and small character sets can uniformly adopt the twin Turing machine indirect computing model and indirect formal method. Thus, Chinese data processing no longer has to rely on small-character-set methods.
\item Existing Chinese word segmentation methods cannot be completely thorough; segmentation at best only approaches "sign set"-like classification, and cannot achieve the thoroughness of "stratified sets".
\end{enumerate}

\begin{figure}
    \centering
    \includegraphics[width=1\linewidth]{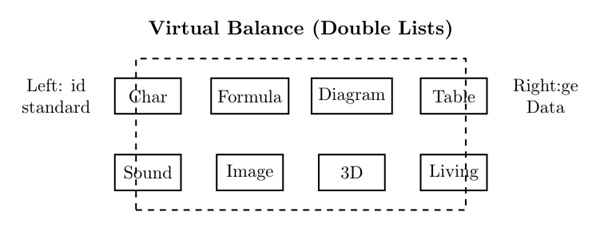}
    \caption{Schematic of indirect formal processing for eight types of data \cite{xmind6}}
    \label{fig:placeholder}
\end{figure}
As shown in Figure 6, eight types of data — characters, formulas, diagrams, tables, sounds, images, three-dimensional objects, living bodies — can all be indirectly formalized based on double lists. This kind of virtual twin Turing machine has three functions: tabularization (double columns), digitization (left column), and word-groupization (right column). In Figure 6, “Wen” refers to generalized text; “Yi” refers to the order-position relations synergistically recorded (understood or expressed) by the twin Turing machine, double lists, and order-position identities.

\section{Conclusion}

The term “twin” in this paper specifically refers to the twin characteristics or essential attributes that arise from combining indirect computation and indirect formalization. It characterizes the essential attribute:
\[
\text{Twin Turing machine virtual computing program} = \text{Good algorithms from indirect computing model} + \text{Optimized data structures from indirect formal method}
\]

The key steps are:
\begin{itemize}
\item From (a) to (b): Using the indirect computing model to further formalize Turing's computable numbers within the target domain, ensuring a series of good algorithms.
\item From (b) to (c): Revealing the connotation of the transformation constraint “N may or may not exist” in the double transformation process of NP-complete problems (deepening understanding from P to NP and simplifying expression from NP to P) — that is, performing indirect computation and indirect formal expression.
\end{itemize}
Among these, adopting an indirect formal method compatible with both large and small strings involves the optimization of data structures. This not only improves Kleene’s string formal theory but also extends beyond small strings to the computer-aided processing of multimedia forms (characters, formulas, diagrams, tables, sounds, images, three-dimensional objects, living bodies) and generalized texts (various types of formal information).

In summary, the indirect computing model can provide optimal paths or best algorithms within the target domain; the indirect formal method can provide optimized data structures within the target domain. Their combination forms a collaborative intelligent computing program for interaction between computer agents (systems) and human subjects (users). Together with the human-computer interface continuously optimized through the support of ontology-based university discipline construction across teaching, learning, research, use, and administration, an ideal collaborative intelligent computing system is obtained. Thus, it can provide a higher level of collaborative intelligent computing for special cases of such systems — cloud computing and cloud-edge computing, as well as distributed, parallel, and grid computing — achieving optimized cloud computing from data centers to knowledge centers.

\section*{Acknowledgments}

First, I thank Academician Lu Ruqian (for supporting and encouraging the publication of my original works) and Academician Zhang Bo (for his attention and for taking time from his busy schedule to talk with me before my trip to Harvard, offering encouragement). Second, I thank Professor Lu Chuan (for supporting, encouraging, and reusing the result data of my two-character Chinese formalization database) and Professor Huang Changning (for inviting me to Microsoft Research Asia in 2002, carefully inquiring and listening to my concept of a Chinese formalization database — the relation database between characters and character groups). Finally, I thank UC Berkeley (for providing me the opportunity to present the design plan of the Chinese formalization database — the relation database between characters and character groups — and its underlying “character-based” theory).

\bibliographystyle{unsrt}

\end{document}